\documentclass[aps,prl,twocolumn,superscriptaddress,groupedaddress]{revtex4}
\usepackage{graphicx}
\usepackage{subfigure}
\usepackage{dcolumn}
\usepackage{bm}
\usepackage{amssymb}
\usepackage{amsmath}
\usepackage{braket}

\begin{document}

%\title{Exactly-solvable number-conserving interacting fermion models with topological superconducting ground states}
\title{Number-conserving interacting fermion models with exact topological superconducting ground states}
\author{Zhiyuan Wang}
\affiliation{School of Physics, Peking University, Beijing 100871, China}
\affiliation{Department of Physics and Astronomy, Rice University, Houston, Texas 77005,
USA}
\author{Youjiang Xu}
\affiliation{Department of Physics and Astronomy, Rice University, Houston, Texas 77005,
USA}
\author{Han Pu}
\affiliation{Department of Physics and Astronomy, Rice University, Houston, Texas 77005,
USA}
\affiliation{Center for Cold Atom Physics, Chinese Academy of Sciences, Wuhan 430071,
China}
\affiliation{Rice Center for Quantum Materials, Rice University, Houston, Texas 77005, USA}
\author{Kaden R.~A. Hazzard}
\affiliation{Department of Physics and Astronomy, Rice University, Houston, Texas 77005,
USA}
\affiliation{Rice Center for Quantum Materials, Rice University, Houston, Texas 77005, USA}
\date{\today}

\begin{abstract}
We present a method to construct number-conserving Hamiltonians whose
ground states exactly reproduce an arbitrarily chosen BCS-type mean-field state.
Such parent Hamiltonians can be constructed not only for the usual $s$-wave
BCS state, but also for more exotic states of this form, including the
ground states of Kitaev wires and 2D topological superconductors. This
method leads to infinite families of locally-interacting fermion models with exact
topological superconducting ground states. After explaining the general technique, we apply
this method to construct two specific classes of models.
The first one is a one-dimensional double wire lattice model with
 Majorana-like degenerate ground states. The second one is a
two-dimensional $p_x+ip_y$ superconducting model, where we also obtain analytic
expressions for topologically degenerate ground states in the presence of
vortices. Our models may provide a deeper conceptual understanding of how
Majorana zero modes could emerge in condensed matter systems, as well as inspire novel
routes to realize them in experiment. 
%Non-Abelian braiding can also be studied in this model in an exact manner.
\end{abstract}

\maketitle

% The following information is for internal review, please remove them for submission
%\widetext
%\leftline{Version 00 as of \today}
%\leftline{Primary authors: Joe E. Physics}
%\leftline{To be submitted to PRL}
%\leftline{Comment to {\tt d0-run2eb-nnn@fnal.gov} by xxx, yyy}
%\centerline{\em D\O\ INTERNAL DOCUMENT -- NOT FOR PUBLIC DISTRIBUTION}

% the following line is for submission, including submission to the arXiv!!
%\hspace{5.2in} \mbox{Fermilab-Pub-04/xxx-E}

%\title{Topological superconductivity in number-conserving models with exactly solvable ground states}

%\email{kaden.hazzard@gmail.com}

%\pacs{71.10.Pm,~03.65.Vf,~74.20.Fg,~74.90.+n}

\paragraph{Introduction.}

%Majorana fermions have sparked interest in condensed matter and cold atoms as emergent quasiparticles with fundamentally new properties, in particular non-Abelian statistics.
Topological superconductors have become an active research area in condensed
matter and cold atom physics~\cite{Kitaev,TPSC-RMP}. From a fundamental
viewpoint, they provide examples of topological phases that have been
classified systematically~\cite{TPorder1,TPorder2,TPorder3,TPorder4,TPorder5,TPorder6}. At a
practical level, the non-Abelian statistics~\cite{MR1991,Nayak1996,Ivanov}
of Majorana zero modes and the robustness of degenerate ground states against
local perturbations have made topological superconductors components of
promising architectures for fault-tolerant quantum computation~\cite{TPQC-RMP}. Experimental signatures~\cite{Exp1,Exp2,Exp3,Exp4} of Majorana
zero modes and topological superconductivity in solid state systems call for more realistic theoretical descriptions of the relevant
physics, for example the effect of interactions.

Most of the theoretical research in this area has begun with
non-interacting mean-field Hamiltonians 
with an effective $p$-wave pairing
term, from which one obtains topologically-protected degenerate ground
states and Majorana zero modes that emerge as effective quasiparticles. In
nature, however, quantum systems typically contain sizable
interacting terms that challenge the validity of mean-field theory. It is
therefore important to understand better the criteria for the aforementioned
topological phenomena to persist in the interacting case. Moreover, the
mean-field approximation breaks the number conservation, which obscures the
connection 
%between the mean-field Majorana operators and the corresponding observables in 
to realistic, number-conserving systems.

Understanding the interplay of topology, number conservation, and interactions is challenging because the interactions usually prevent exact solution by analytic or numerical techniques.
%Resolving these issues is challenging  due to the importance of interactions. 
In one dimension, special
tools are available, and progress has been made using bosonization~\cite{Bos1,Bos2,Bos3} and numerical methods [density-matrix
renormalization group~(DMRG)]~\cite{PZoller}. Exactly solvable models in this
area are still rare~\cite{InteractingKitaev}, and the three number-conserving Majorana models~\cite{RGK,SDiehl,NLang} that have been proposed are
one-dimensional~(1D). Having new families of exactly solvable models
with realistic local interactions, especially in higher dimensions, will
therefore shed light on the characterization of topological phenomena and
Majorana zero modes in intrinsically interacting and number-conserving
systems. In addition, these results provide new Hamiltonians that can be used to experimentally realize topological states.

In this letter we take a bottom-up approach to these fundamental issues: Starting from
a general BCS-type mean-field ground state $\ket{G}$, we show how to
construct number-conserving parent Hamiltonians that have $\ket{G}$ as a
ground state (with no approximation). This construction enables us to
realize the physics of Majorana zero modes in interacting number-conserving
systems in an exact manner. Following the general construction, we build
specific models including a 1D Majorana double wire and a 2D $p_x+ip_y$
topological superconductor, and obtain analytic expressions for the
degenerate ground states in the presence of edges and vortices.

\paragraph{General construction.}

Suppose we have an effective mean-field Hamiltonian in some BCS-like theory
in any dimension with or without spin 
\begin{eqnarray}  \label{eq:Kmf}
K_{\mathrm{mf}}&=&\sum_{p}[\xi_{p}a^\dagger_{p}a_{p}+\frac{1}{2}%
(\Delta^*_{p}a_{\bar{p}}a_{p}+\Delta_{p}a^\dagger_{p}a^\dagger_{\bar{p}})] 
\notag \\
&=&\sum_pE_p\alpha_p^\dagger\alpha_p+\mathrm{const},
\end{eqnarray}
where $E_{p}=\sqrt{\xi_p^2+|\Delta_p|^2}$, $p$ indexes the single-particle states (including momentum, spin, or any other quantities necessary), $\bar{p}$ denotes the time
reversed state of $p$, and $\alpha_p=u_p a_p-v_p a_{\bar{p}}^\dagger$ are
Bogoliubov quasi-particle operators with $|u_p|^2+|v_p|^2=1$ and $%
v_p/u_p=-(E_p-\xi_p)/\Delta_p^*$. The BCS-like ground state of $K_{\text{mf}}
$ is~(up to normalization) 
\begin{equation}  \label{eq:G_BCS}
|G\rangle=\sideset{}{'}\prod_{p} \alpha_{p}\alpha_{\bar{p}}|0\rangle,
\end{equation}
where the prime means each pair $p\bar{p}$ appears exactly once.

Our goal is to construct a number-conserving Hamiltonian whose ground state
is $|G\rangle $. To do this, we first separate each $\alpha_{p}$ into
creation and annihilation parts $\alpha_{p}\equiv C_{p}-S_{p}^{\dagger }$
with $C_{p}=u_{p}a_{p}$ and $S_{p}^{\dagger }=v_{p}a_{\bar{p}}^{\dagger }$,
and define 
\begin{equation}
\label{eq:A_pp}
\hat{A}_{pp'}=S^\dagger_{p}\alpha_{p'}+S^\dagger_{p'}\alpha_p.%=S^\dagger_{p}C_{p'}+S^\dagger_{p'}C_p.   
\end{equation}%
From $\alpha_p|G\rangle=0$, we know that $\hat{A}_{pp^{\prime
}}|G\rangle =0$, and thus a parent Hamiltonian for $|G\rangle $ can be
constructed by 
\begin{equation}
\hat{H}=\sum_{p_{1}p_{2}p_{3}p_{4}}H_{p_{1}p_{2};p_{3}p_{4}}\hat{A}%
_{p_{1}p_{2}}^{\dagger }\hat{A}_{p_{3}p_{4}},  \label{eq:Hparen}
\end{equation}%
where the matrix $H_{p_{1}p_{2};p_{3}p_{4}}$ is required to be Hermitian.
This construction suffices for $\ket{G}$ to be a zero-energy eigenstate of $%
\hat{H}$. To ensure it is a ground state, we require that the matrix $%
H_{p_{1}p_{2};p_{3}p_{4}}$ is positive-definite. Notice that $\hat{H}$
conserves total particle number ${\hat{N}}=\sum_{p}a_{p}^{\dagger }a_{p}$
since ${\hat{A}}_{pp^{\prime }}$ can be rewritten as 
\begin{equation}
{\hat{A}}_{pp^{\prime }}=S_{p}^{\dagger }C_{p^{\prime }}+S_{p^{\prime
}}^{\dagger }C_{p},  \label{eq:A_pp2}
\end{equation}%
which follows because $S_{p}^{\dagger }S_{p^{\prime }}^{\dagger
}+S_{p^{\prime }}^{\dagger }S_{p}^{\dagger }$ vanishes by fermionic
antisymmetry. Then the ground state of $\hat{H}$ with a definite particle
number $N$ is simply given by the projection of $|G\rangle $ to the $N$%
-particle subspace $|G_{N}\rangle =\hat{P}_{N}|G\rangle $.

\paragraph{The double wire model.}

As a first specific example of this construction, we construct
one-dimensional models that reproduce the ground states of Kitaev's 1D wire~\cite{Kitaev}, $\hat{H}_{\mathrm{Kitaev}}=\sum_{j}(-tc_{j}^{%
\dagger }c_{j+1}+\Delta c_{j}c_{j+1}+\mathrm{H.c.})-\mu \hat{N}$, where $t$, 
$\mu $, and $\Delta $ denote the hopping amplitude, the chemical potential
and the superconducting gap, respectively. The Kitaev's model has a special point at $\mu =0,~\Delta =t$ where the Hamiltonian can be rewritten as a sum of mutually commuting local operators and the spectrum is non-dispersing~\cite{Kitaev}. To simplify our calculation, we focus on this special point~(though our construction protocol is general and could be applied to other points as well~\cite{Suppl}) where the Hamiltonian has doubly degenerate ground states given by~(up to normalization) 
\begin{equation}\label{eq:KitaevG}
%\ket{G_e} =\prod_{k}\alpha_k|0\rangle,~\ket{G_o}= c^\dagger_0\prod_{k}\alpha_k\ket{0}.
\ket{G^e} =\exp\left(\sum_{i<j}c^\dagger_ic^\dagger_j\right)|0\rangle,~\ket{G^o}= \tilde{c}^\dagger_{k=0}|G^e\rangle,
\end{equation}
where $\tilde{c}^\dagger_{k}=\frac{1}{\sqrt{L}}\sum^L_{j=1}e^{ikj} c^\dagger_{j}$~\cite{Suppl}. The superscripts $e,o$ denote even
and odd fermion parity, respectively.

We consider a double wire geometry that has two parallel one-dimensional
chains, with fermion creation operators on each chain given by $%
a^\dagger_j\equiv c^\dagger_{j,1},~b^\dagger_j\equiv
c^\dagger_{j,2}$, respectively. Our aim is to construct a
number-conserving lattice model on these two wires whose ground states are
direct products of the Kitaev ground states on each wire, projected to fixed
total particle number $\ket{G_N}={\hat P}_N(\ket{G_A}\otimes\ket{G_B})$. We
will show that the resulting Hamiltonian also leads to Majorana-like edge
modes and robust ground state degeneracy.

The direct product of Kitaev ground states $|G_A\rangle\otimes |G_B\rangle$
is annihilated by Bogoliubov operators~\cite{Suppl} 
\begin{eqnarray}  \label{eq:alpha_k}
\alpha_{k\sigma}&=&\frac{e^{i\frac{k}{2}}}{\sqrt{2}}\left(-\sin\frac{k}{2}%
\tilde{c}_{k\sigma}+i\cos\frac{k}{2}\tilde{c}^\dagger_{-k\sigma}\right)-(k%
\to-k)  \notag \\
&\equiv& C_{k\sigma}-S^\dagger_{k\sigma},
\end{eqnarray}
where $\sigma=1,2$, and the quasimomentum $k$ is quantized
with open boundary condition $k=\frac{m\pi}{L},~m=1\ldots(L-1)$~\footnote{%
Notice that with open boundary condition $k=m\pi/L$, the $\tilde{c}%
_{k\sigma},\tilde{c}^\dagger_{k^{\prime }\sigma^{\prime }}$ no longer
satisfy the canonical anti-commutation relations, but $\{\alpha_{k\sigma},%
\alpha^\dagger_{k^{\prime }\sigma^{\prime }}\}=\delta_{kk^{\prime
}}\delta_{\sigma\sigma^{\prime }}$.}.

\begin{figure}[tbp]
\includegraphics[width=0.7\linewidth]{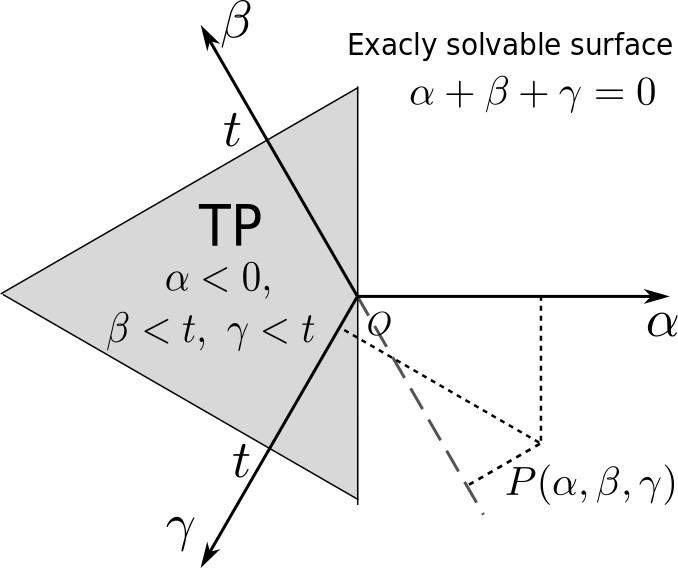}
\caption{Topological phase~(TP) of the double wire system. The constraint $%
\protect\alpha +\protect\beta +\protect\gamma =0$ restricts the parameter
space to the 2D plane drawn in this figure, where the coordinate $(\protect%
\alpha ,\protect\beta ,\protect\gamma )$ of an arbitrary point $P$ is given
by the projections from $P$ to the $\protect\alpha ,\protect\beta ,\protect%
\gamma $ axes. The conditions $\protect\alpha <0,\protect\beta <t,\protect%
\gamma <t$, which guarantee $\hat{H}$ to be positive definite, give a
triangle region~(shaded area) in this plane.}
\label{fig:triangle}
\end{figure}

Having identified the $\alpha_{k\sigma }$, and therefore the $C_{k\sigma }$
and $S_{k\sigma }^{\dagger }$, Eq.~\eqref{eq:Hparen} gives a family of
parent Hamiltonians. By choosing 
\begin{gather}
H_{p_{1}p_{2};p_{3}p_{4}}=\frac{16}{L^{2}}\sum_{j=1}^{L}\sin (k_{1}j)\sin
(k_{2}j)\sin (k_{3}j)\sin (k_{4}j)  \notag \\
\times \left[ p\delta_{\sigma_{1}\sigma_{2}\sigma_{3}\sigma
_{4}}+q\delta_{\sigma_{1}\sigma_{2}}\delta_{\sigma_{3}\sigma
_{4}}-r(\delta_{\sigma_{1}\sigma_{3}}\delta_{\sigma_{2}\sigma
_{4}}+\delta_{\sigma_{1}\sigma_{4}}\delta_{\sigma_{2}\sigma_{3}})%
\right] ,  \label{eq:dbwireHpp}
\end{gather}%
where $p_j=(k_j,\sigma_j)$ and $p,q,r$ are arbitrary real
coefficients, one obtains Hamiltonians ${\hat{H}}$ with interactions that
are local in real space~\cite{Suppl}, 
\begin{eqnarray}\label{eq:dbwireH}
\hat{H}&=&\sum^{L-1}_{j=1}\left\{-t\left[(a^\dagger_ja_{j+1}+b^\dagger_jb_{j+1}+\mathrm{H.c.})-1\right.\right.\nonumber\\
&&\left.+2(n^a_j-\frac{1}{2})(n^a_{j+1}-\frac{1}{2})+2(n^b_j-\frac{1}{2})(n^b_{j+1}-\frac{1}{2})\right]\nonumber\\
&&\left.-(\alpha J^\dagger_{||,j}J_{||,j}+\beta J^\dagger_{=,j} J_{=,j}+\gamma J^\dagger_{\times,j} J_{\times,j})\right\},
\end{eqnarray}
where $J_{||,j}=b_{j}a_{j}-b_{j+1}a_{j+1},$ $%
J_{=,j}=a_{j+1}a_{j}-b_{j+1}b_{j},~J_{\times ,j}=b_{j+1}a_{j}-b_{j}a_{j+1}$
and $\alpha ,\beta ,\gamma $ are real numbers determined by $p,q,r$~\cite{Suppl} with constraint $\alpha +\beta +\gamma =0$. In order for the matrix
Eq.~(\ref{eq:dbwireHpp}) to be positive-definite, the $\alpha, \beta, \gamma 
$ should satisfy $\alpha <0,\beta <t,\gamma <t$, which, combined with $%
\alpha +\beta +\gamma =0$, gives the triangle region that is shown in Fig.~\ref{fig:triangle}. The center of the triangle $\alpha=-t,~\beta=\gamma=t/2$ reproduces the model in Ref.~\cite{SDiehl}.
%~\footnote{The relation between $t,\alpha,\beta,\gamma$ and $p,q,r$ is $p=2(t+\alpha-\gamma),~q=\gamma-\beta,r=\alpha$.}

Since $\hat{H}$ preserves total particle number $N=N_{A}+N_{B}$ and single
wire fermion parity $P^{A,B}=(-1)^{\hat{N}_{A,B}}$, ground states in each $N$-particle sector are doubly degenerate. For example, if $N$ is even, we
have~(up to normalization) 
\begin{equation}
|G_{N}^{ee}\rangle =\left[ \sum_{i<j}(a_{i}^{\dagger }a_{j}^{\dagger
}+b_{i}^{\dagger }b_{j}^{\dagger })\right] ^{\frac{N}{2}}|0\rangle
,~|G_{N}^{oo}\rangle =\tilde{a}_{0}^{\dagger }\tilde{b}_{0}^{\dagger
}|G_{N-2}^{ee}\rangle .
\end{equation}%
This degeneracy is topologically protected in the sense that all local perturbations in the bulk, even including the ones that
violate single wire parity, take the form of an identity matrix when projected to the ground state subspace. For example, using the same arguments as in Refs.~\cite{NLang,SDiehl}, we explicitly find that the energy splitting $\Delta E$ due to perturbation $a_{j}^{\dagger }b_{j}+\mathrm{H.c.}$ scales
as $\Delta E\sim e^{-j/l_{0}}$~(assuming $j<L/2$) for some finite length
scale $l_{0}$.

\paragraph{The 2D $p_x+ip_y$ model.}
Majorana zero modes are expected to appear
at certain boundaries and in the cores of vortices 
 in $p_{x}+ip_{y}$ superconductors. We now show that Eq.~(%
\ref{eq:Hparen}) can be used to construct number-conserving parent
Hamiltonians for $p_{x}+ip_{y}$ topological superconductors that share the
same ground state as the mean-field model 
\begin{equation}
\hat{K}=\int_{S}\left[ \frac{\nabla \psi_{z}^{\dagger }\cdot \nabla \psi
_{z}}{2m}-(\Delta \psi_{z}\partial_{\bar{z}}\psi_{z}+\mathrm{H.c.})+\mu
\psi_{z}\psi_{z}^{\dagger }\right] d^{2}z,  \label{eq:K_2Dpp}
\end{equation}%
where $S$ denotes an arbitrary region in the 2D plane with complex
coordinates $z=x+iy$, $\partial_{z}=(\partial_{x}-i\partial_{y})/2$, $%
\partial_{\bar{z}}=(\partial_{x}+i\partial_{y})/2$, $d^{2}z=dxdy$, and $%
\psi_{z}$ is the fermionic annihilation operator at position $z$. The term $%
\Delta \psi_{z}\partial_{\bar{z}}\psi_{z}+\mathrm{H.c.}$ characterizes
chiral $p$-wave pairing, and $\mu \psi_{z}\psi_{z}^{\dagger }$ is the
chemical potential term~(which differs from the usual convention by a
constant). Although in principle we can construct number-conserving parent
Hamiltonians for all values of $(m,\Delta ,\mu )$, in the following we only
consider a special point $\mu =\frac{m\Delta ^{2}}{2}$ and use natural units 
$2m=m\Delta =1$ for simplicity. With an integration by parts, $\hat{K}$ can
be separated into a bulk Hamiltonian and a boundary term $\hat{K}=\hat{K}_{%
\mathrm{bulk}}+\hat{K}_{\mathrm{bound}}$, with 
\begin{eqnarray}
\hat{K}_{\mathrm{bulk}} &=&\int_{S}(2\partial_{z}\psi_{z}^{\dagger }-\psi
_{z})(2\partial_{\bar{z}}\psi_{z}-\psi_{z}^{\dagger })d^{2}z,  \notag
\label{eq:K_bb} \\
\hat{K}_{\mathrm{bound}} &=&-i\oint_{\partial S}(\psi_{z}^{\dagger
}\partial_{z}\psi_{z}dz+\psi_{z}^{\dagger }\partial_{\bar{z}}\psi_{z}d%
\bar{z}),
\end{eqnarray}%
where $\partial S$ denotes the boundary of $S$. Since $\hat{K}_{\mathrm{bulk}}$ is by construction positive-definite, the ground states of $\hat{K}$
should be annihilated by the operator $\alpha_{z}\equiv 2\partial_{\bar{z}%
}\psi_{z}-\psi_{z}^{\dagger }$ for all $z\in S$ in order to minimize $\hat{%
K}_{\mathrm{bulk}}$~(we will account for the boundary term momentarily). The
ground states with even fermion parity can in general be constructed as 
\begin{equation}
|G^{e}\rangle =\exp \left[ \frac{1}{2}\int_{S}g(z,z^{\prime })\psi
_{z}^{\dagger }\psi_{z^{\prime }}^{\dagger }d^{2}zd^{2}z^{\prime }\right]
|0\rangle ,  \label{eq:Gg}
\end{equation}%
where the two-particle wave function $g(z,z^{\prime })$ satisfies $%
g(z,z^{\prime })=-g(z^{\prime },z)$ and 
\begin{equation}
2\partial_{\bar{z}}g(z,z') = \delta^{2}(z-z'),  \label{eq:gzz}
\end{equation}%
which guarantees that $\alpha_{z}|G^{e}\rangle =0$. To simultaneously
minimize the boundary term $\hat{K}_{\mathrm{bound}}$, the function $%
g(z,z^{\prime })$ should satisfy certain boundary conditions that depend on
the geometry of the region $S$, which we will discuss later.

To find a parent Hamiltonian for the mean-field ground state $|G^e\rangle$,
we again follow our general construction given in Eqs.~(\ref{eq:Hparen}) and
(\ref{eq:A_pp2}) where we identify $C_z=2\partial_{\bar{z}} \psi_z$ and $%
S^\dagger_z=\psi^\dagger_z$, leading to 
\begin{equation}  \label{eq:Hparen_pos}
\hat{H}_{\mathrm{bulk}}=\int_S W(z_1,z_2;z_3,z_4)A^\dagger_{z_1, z_2}A_{z_3,
z_4}\prod^4_{j=1}d^2z_j,
\end{equation}
where $W(z_1,z_2;z_3,z_4)$ is a positive-definite Hermitian matrix. We 
further restrict ourselves to 
Hamiltonians describing short-ranged interactions, i.e. $%
W(z_1,z_2;z_3,z_4)$ tends to zero sufficiently fast when the distance
between any two points $|z_i-z_j|$   becomes large.
Furthermore, the boundary term $\hat{K}_{\mathrm{bound}}$ in Eq.~(\ref{eq:K_bb}) should be added into $\hat{H}_{\mathrm{bulk}}$ to uniquely pick
out the same set of ground states as $\hat{K}$%
%in order to eliminate unwanted degeneracy and stabilize the topological phase~\cite{Suppl}
\begin{equation}  \label{eq:H_pip}
\hat{H}=\hat{H}_{\mathrm{bulk}}+\hat{K}_{\mathrm{bound}}.
\end{equation}
The new interacting Hamiltonian $\hat{H}$
harbors the topological $p_x+ip_y$ ground state. It is 
number-conserving because both $\hat{H}_{\mathrm{bulk}}$ and $\hat{K}_{%
\mathrm{bound}}$ preserve total particle number.

As a specific example, we choose $W(z_1,z_2;z_3,z_4)=w(z_1-z_3)%
\delta^2(z_1-z_2)\delta^2(z_3-z_4)/4$ in Eq.~(\ref{eq:Hparen_pos}) where $%
w(-z)=[w(z)]^*$ and $w(0)=1$,
after rearranging terms we get 
\begin{eqnarray}
\hat{H}&=&\int_S\nabla\psi^\dagger_{z}\cdot\nabla \psi_{z}d^2 z \\
&+&4\int_S w(z-z')\psi^{\dagger}_{z}(\partial_{z^{\prime
}}\psi^\dagger_{z^{\prime }})\psi_{z^{\prime }}\partial_{\bar{z}}\psi_z d^2z
d^2z^{\prime }.  \notag
\end{eqnarray}
The function $w(z-z^{\prime })$ is required to be positive-definite in the
matrix sense~(i.e. $\tilde{w}_{\mathbf{k}}=\int w(\mathbf{r})e^{i\mathbf{k}%
\cdot\mathbf{r}}d^2\mathbf{r}$ should be real and positive for all $\mathbf{k%
}$), and should decay sufficiently fast as $|z-z^{\prime }|$ becomes large. 
%The Hamiltonian $H_{\mathrm{bulk}}$ constructed in Eq.~(\ref{eq:Hparen_pos}) is the parent Hamiltonian of the ground states of $K_{\mathrm{bulk}}$, which has the subtle problem that in open boundary conditions, chiral edge modes with nonzero momentum are found to have exactly zero energy, leading to unwanted  degeneracy. We should therefore add 

\paragraph{Degenerate ground states with vortices.}
It is known that vortices in the mean-field model Eq.~(\ref{eq:K_2Dpp}) have localized Majorana zero-modes~\cite{NRead,Ivanov} giving rise to topologically-protected degeneracy and non-Abelian statistics. It would be interesting to see whether these important properties survive in our number-conserving model, as these properties are crucial for the realization of topological quantum computation~\cite{TPQC-RMP}. As a first step, we need to obtain the ground state wavefunctions, a calculation whose results we present in this section. 
One remarkable feature of our model is that the analytic expressions of
degenerate ground states could be exactly obtained even when there are an
arbitrary number of vortices in the 2D plane. 
These explicit expressions could give us deeper insight into the topological properties of the ground states and 
provide a platform to study non-Abelian statistics in number-conserving interacting models in an exact manner. 
%To study Majorana zero modes in our number-conserving $p_x+ip_y$ model, we have to include edges and vortices in the 2D plane, since Majorana modes appear in the transition region~(domain walls) between the topological superconducting phase and vacuum~\cite{NRead}. 

\begin{figure}[tbp]
\includegraphics[width=0.6\linewidth]{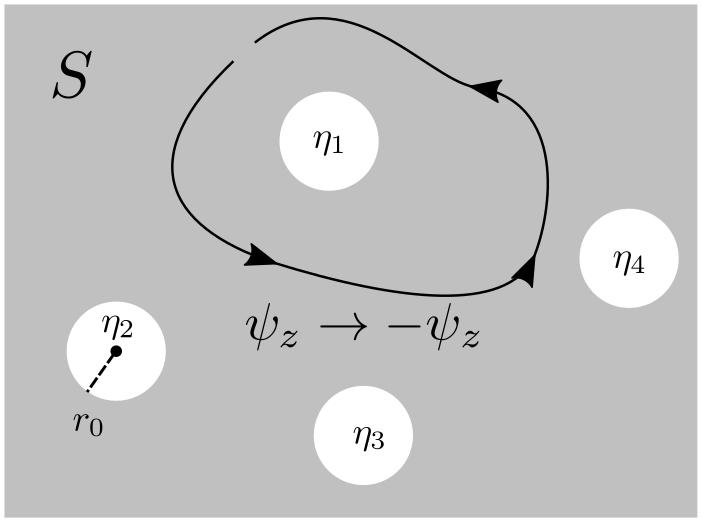}
\caption{Topological superconducting phase in an unbounded 2D plane with $2M$
vortices located at $\protect\eta_1,\protect\eta_2,\ldots, \protect\eta_{2M}$. In this figure we show the $2M=4$ case. In our gauge convention fermion
fields acquire a minus sign on going around each vortex.}
\label{fig:holes}
\end{figure}

We consider the geometry shown in Fig.~\ref{fig:holes}, with $2M$ vortices
lying in an unbounded 2D plane located at $\eta_1,\eta_2,\ldots ,\eta_{2M}$,
respectively. We assume that the 
%vacuum region in the 
core of each vortex is
%separated from the topological superconducting phase by a circular domain
%wall 
localized inside a radius $r_0$ much smaller than the 
minimal distance between any two vortices. We use the gauge convention in
which the superconducting order parameter is the same everywhere [i.e. still
consider the same Hamiltonians in Eqs.~(\ref{eq:K_2Dpp}) and (\ref{eq:Hparen_pos})] while fermion fields are anti-periodic around each vortex $%
\psi^\dagger_{\theta+2\pi}=-\psi^\dagger_{\theta}$.

In the mean-field model Eq.~(\ref{eq:K_2Dpp}) there is a Majorana zero-mode $\gamma_j$ with $%
\gamma_j^\dagger=\gamma_j$ and $[\gamma_j,\hat{K}]=0$ localized at the $j^{%
\mathrm{th}}$ vortex. In total we have $2M$ localized Majorana modes $%
\gamma_1,\gamma_2,\ldots, \gamma_{2M}$ which could be combined to $M$
independent fermion operators $a_j=(\gamma_{2j-1}+i\gamma_{2j})/2$, leading
to $2^M$ degenerate mean-field ground states $|n_1,n_2,\ldots, n_M\rangle$
with $n_j=0,1,~1\leq j\leq M$. In our number-conserving model defined in
Eq.~(\ref{eq:H_pip}), the ground states with $N$ particles are obtained by
projecting the mean-field ground states to the $N$-particles sector. Only
those mean-field states with fermion parity equal to $(-1)^N$ survive this
projection, therefore we are left with $2^{M-1}$ fold degeneracy in each
sector.

As an example, we consider the $2M=4$ case and assume $N$ to be even. One of
the mean-field ground states $|G_{12,34}\rangle$ could be constructed by
Eq.~(\ref{eq:Gg}) with 
\begin{eqnarray}  \label{eq:g_1234}
g_{12,34}(z,z^{\prime })&=&\left[\sqrt{\frac{(z-\eta_1)(z-\eta_2)(z^{\prime
}-\eta_3)(z^{\prime }-\eta_4)}{(z^{\prime }-\eta_1)(z^{\prime
}-\eta_2)(z-\eta_3)(z-\eta_4)}}\right.  \notag \\
&&\left.+(z\leftrightarrow z^{\prime })\vphantom{\frac{1}{2}}\right]\frac{1}{4\pi(z-z^{\prime })},
\end{eqnarray}
where the terms in the bracket is to guarantee that $g_{12,34}(z,z^{\prime })
$ is anti-periodic around each vortex, in accordance with our gauge
convention. From the identity $2\partial_{\bar{z}}\frac{1}{z-z^{\prime }}%
=2\pi\delta^2(z-z^{\prime })$ it is easy to see that $g_{12,34}(z,z^{\prime
})$ satisfies Eq.~(\ref{eq:gzz}), and it can be checked that the state $%
|G_{12,34}\rangle$ also minimizes $\hat{K}_{\mathrm{bound}}$ up to some
small corrections~\cite{2D_TPSC_mf}. Applying the projection operator $\hat{P%
}_N$, we get an $N$-particle ground state with multiparticle wave
function~(up to normalization) 
\begin{equation}  \label{eq:MR}
\psi_{12,34}(z_1,z_2\ldots z_N)=\mathrm{Pf}\{g_{12,34}(z_i,z_j)\},
\end{equation}
where Pf denotes the Pfaffian of the anti-symmetric $N\times N$ matrix $%
g_{12,34}(z_i,z_j)$. %\begin{equation}
%\mathrm{Pf}\{M_{ij}\}=\frac{1}{2^{N/2}(N/2)!}\sum_{\sigma}\mathrm{sgn}(\sigma)\prod^{N/2}_{j=1}M_{\sigma(2j-1),\sigma(2j)}.
%\end{equation} 
The form of the wave function given in Eq.~(\ref{eq:MR}) is very similar to
one of the Moore-Read Pfaffian states with four quasiholes~\cite{MR1991,Nayak1996}, which were constructed to describe the excitations in
the $\nu=5/2$ fractional quantum hall effect~(FQHE). By permuting the
indices $1,2,3,4$ we get two other degenerate ground states with wave
functions $\psi_{13,24}(z_1\ldots z_N)$ and $\psi_{14,23}(z_1\ldots z_N)$.
However, using the same method in Ref.~\cite{Nayak1996} we can prove that
these three states are linearly dependent and the space spanned by them is
actually two dimensional, consistent with our previous argument. 
%The projected state $P_N|G_a\rangle$ (where $a=(12,34)$ or $(13,24)$) can be shown~\cite{NRead} to be a Pfaffian state, with multi-electron wave function $\psi(w_1,w_2\ldots w_N)=\mathrm{Pf}[g(w_i-w_j)]$, where $w_j=x_j+iy_j$ denotes the position of $j$-th electron. 

The non-Abelian statistics of the mean-field ground states of the $p_x+ip_y$
model have been well-studied in Ref.~\cite{Ivanov}. Braiding the $j^{\mathrm{%
th}}$ and the $(j+1)^{\mathrm{th}}$ vortices adiabatically gives rise to a
unitary rotation $\hat{B}_{j j+1}=\exp(\frac{\pi}{4}\gamma_{j+1}\gamma_j)$
on the ground state subspace. In our number-conserving model constructed in
Eq.~(\ref{eq:H_pip}), the process of braiding preserves total particle
number, thus we have to recalculate the Berry's matrix for each $N$-particle
sector, which will be the subject of future work. We expect that such
calculation could be done using similar methods in Ref.~\cite{Bonderson2011}%
, where unitary evolution of the Moore-Read Pfaffian states due to braiding
of quasiholes are calculated with the help of the plasma analogy.

\paragraph{Conclusion.}

We have constructed infinite families of number-conserving, interacting
Hamiltonians with exact BCS-like ground states, with specific
models including a 1D Majorana double wire and a 2D $p_x+ip_y$ topological
superconductor. In the $p_x+ip_y$ model we obtained analytic expressions of
degenerate ground states with four vortices, and pointed out their
similarity to the Moore-Read Pfaffian states with four quasiholes
constructed in the $\nu=5/2$ FQHE context. Our models give us a deeper
theoretical understanding of topological phenomena in interacting systems,
set a viable framework for building more realistic models of topological
superconductors, and may provide useful guidelines for experimental
realization of Majorana zero modes. \acknowledgements
We thank Matthew Foster and Bhuvanesh Sundar for discussions. HP was supported by the NSF and the
Welch Foundation (Grant No. C-1669). KRAH was supported in part with funds
from the Welch Foundation (Grant No. C-1872).

\clearpage
\newpage

\clearpage 
\setcounter{equation}{0}%
\setcounter{figure}{0}%
\setcounter{table}{0}%
\renewcommand{\thetable}{S\arabic{table}}
\renewcommand{\theequation}{S\arabic{equation}}
\renewcommand{\thefigure}{S\arabic{figure}}

\onecolumngrid

\begin{center}
{\Large Supplemental Material to \\ ``Number-conserving interacting fermion models with exact topological
superconducting ground states'' }

\vspace*{0.5cm}

Zhiyuan Wang, Youjiang Xu, Han Pu, and Kaden R. A. Hazzard

\vspace{1cm}

\end{center}

\section{The Double wire model}
In this section we first present the detailed derivations of Eqs.~(\ref{eq:KitaevG}-\ref{eq:dbwireH}) in the main text, and then we discuss some alternative derivations of the parent Hamiltonian.  
\subsection{Diagonalization of Kitaev's Hamiltonian in momentum space with open boundary}
To obtain the Bogoliubov operators in Eq.~(\ref{eq:alpha_k}) and the form of ground states in Eq.~(\ref{eq:KitaevG}) in our main text, here we present the momentum space diagonalization of Kitaev's Hamiltonian with $\Delta=t$
\begin{equation}
H_{\mathrm{Kitaev}}=\sum_{j} t(-c^\dagger_{j}c_{j+1}+ c_{j}c_{j+1}+\mathrm{H.c.})-\mu \hat{N}.
\end{equation}
To this end we search for Bogoliubov eigenmodes defined as
\begin{equation}
\alpha_k=\sum^L_{j=1}(u^k_{j} c_j+v^k_{j} c^\dagger_j).
\end{equation}
Being the eigenmodes of $H_{\mathrm{Kitaev}}$ with energy $E_k>0$, they satisfy $[\alpha_k,H_{\mathrm{Kitaev}}]= E_k\alpha_k$, which gives difference equations on $u^k_{j},v^k_{j}$
\begin{eqnarray}\label{eq:recursive}
(E_k+\mu)u^k_j&=&-t(u^k_{j-1}+u^k_{j+1})+t(v^k_{j-1}-v^k_{j+1}),\nonumber\\
(E_k-\mu)v^k_j&=&t(v^k_{j-1}+v^k_{j+1})+t(u^k_{j+1}-u^k_{j-1}),
\end{eqnarray}
with boundary conditions 
\begin{eqnarray}\label{eq:obc}
u^k_0=v^k_0,~u^k_{L+1}=-v^k_{L+1}.
\end{eqnarray}
To solve these equations, we notice that the ansatz solutions
\begin{eqnarray}
u^k_j=\lambda(k) e^{i(kj-\theta_k)}-\lambda(-k) e^{-i(kj-\theta_k)},~~v^k_j=e^{i(kj-\theta_k)}-e^{-i(kj-\theta_k)}
\end{eqnarray}
with $\lambda(k)=-i\frac{2\Delta\sin k}{E_k+\mu+2t\cos k}$ and $E_k=\sqrt{\mu^2+4t\mu\cos k+4t^2}$ satisfy Eq.~(\ref{eq:recursive}). The boundary conditions in Eq.~(\ref{eq:obc}) give constraints on the quasi-momentum $k$ and the real parameter $\theta_k$
\begin{eqnarray}\label{eq:sol1}
k(L+1)=2\theta_k+m\pi,~~m\in \mathbf{Z},~~~\tan\theta_k=\frac{2\Delta\sin k}{E_k+\mu+2t\cos k},
\end{eqnarray}
where $0< k< \pi$ and $0<\theta_k< \pi/2$. The Bogoliubov operators are~(up to normalization)
\begin{equation}\label{eq:alpha_kS}
\alpha_k=e^{i\theta_k}(i\cos\theta_k c^\dagger_{-k}-\sin\theta_k c_k)-(k\to -k).%\sum_j[\tan\theta_k\cos(kj-\theta_k) c_j-\sin(kj-\theta_k)c^\dagger_j].
\end{equation}
As an aside, we mention that if we replace Eq.~(\ref{eq:alpha_k}) in our main text by Eq.~(\ref{eq:alpha_kS}) and use the same matrix in Eq.~(\ref{eq:dbwireHpp}), then, still following our general construction, we can get a bigger family of number-conserving, short-range interacting Hamiltonians with ground states $|G^{e,o}\rangle$ depending on $\mu/t$, and this method can be generalized to construct parent Hamiltonians for Kitaev's ground states at arbitrary points $(t,\Delta,\mu)$~(even including points in the topologically trivial phase).  

At the $\mu=0$ point we get especially simple expressions
\begin{equation}
E_k=2t,~~\theta_k=\frac{k}{2},~~k=\frac{m\pi}{L},~m=1\ldots(L-1),
\end{equation}
which leads to the single wire version of Bogoliubov operators in Eq.~(\ref{eq:alpha_k}) after normalization. To verify that the expressions given in Eq.~(\ref{eq:KitaevG}) are indeed the ground states of $H_{\mathrm{Kitaev}}$, we show that $|G^e\rangle$ and $|G^o\rangle$ are annihilated by all $\alpha_k$. We have
\begin{equation}
c_k|G^e\rangle=\frac{1}{\sqrt{L}}\sum^L_{j=1}e^{-ikj}c_j~\exp\{\sum_{i<j'}c^\dagger_{i}c^\dagger_{j'}\}|0\rangle=\frac{1}{\sqrt{L}}\sum_{j,j'}e^{-ikj}\mathrm{sgn}(j'-j)c^\dagger_{j'}|G^e\rangle=[i\cot\frac{k}{2}c^\dagger_{-k}-\frac{1+(-1)^m}{1-e^{ik}}c^\dagger_{k=0}]|G^e\rangle,
\end{equation}
where $\mathrm{sgn}(x)=x/|x|$ for $x\neq 0$ and $\mathrm{sgn}(0)=0$. It follows that
\begin{equation}
[e^{i\frac{k}{2}}\sin\frac{k}{2}c_k-(k\to-k)]|G^e\rangle=[e^{i\frac{k}{2}}i\cos\frac{k}{2}c^\dagger_{-k}-(k\to-k)]|G^e\rangle,
\end{equation}
leading to $\alpha_k|G^e\rangle=0$. Furthermore, it can be easily checked that $\{c^\dagger_{k=0},\alpha_k\}=0$ for all $k=m\pi/L,~1\leq m\leq L-1$, thus $\alpha_k|G^o\rangle=\alpha_kc^\dagger_{k=0}|G^e\rangle=-c^\dagger_{k=0}\alpha_k|G^e\rangle=0$. We conclude that Eq.~(\ref{eq:KitaevG}) indeed gives us the ground states of $H_{\mathrm{Kitaev}}$ at the point $t=\Delta,~\mu=0$.

\subsection{Detailed derivation of Eq.~(\ref{eq:dbwireH})}
To verify that the combination of Eqs.~(\ref{eq:Hparen}) and (\ref{eq:dbwireHpp}) indeed give the local form of Eq.~(\ref{eq:dbwireH}), we first notice that 
\begin{eqnarray}
\alpha_{k\sigma}&\equiv& C_{k\sigma}-S^\dagger_{k\sigma}=\frac{e^{i\frac{k}{2}}}{\sqrt{2}}\left(i\cos\frac{k}{2}c^\dagger_{-k,\sigma}-\sin\frac{k}{2}c_{k\sigma}\right)-(k\to-k)\nonumber\\
&=&\frac{1}{\sqrt{2L}}\sum^{L-1}_{j=1}\sin kj (c_{j+1,\sigma}+c^\dagger_{j+1,\sigma}-c_{j\sigma}+c^\dagger_{j\sigma}),~~k=\frac{m\pi}{L},~m=1\ldots(L-1).
\end{eqnarray}
Using the completeness and orthonormality of $\sin kj$,
\begin{equation}
\sum_{k}\sin kj \sin kj'=\frac{L}{2}\delta_{jj'},
\end{equation}
we have 
\begin{eqnarray}\label{eq:S_j}
\frac{4}{\sqrt{2L}}\sum_k (C_{k\sigma}-S^\dagger_{k\sigma})\sin kj =c_{j+1,\sigma}+c^\dagger_{j+1,\sigma}-c_{j\sigma}+c^\dagger_{j\sigma}\equiv C_{j\sigma}-S_{j\sigma}^\dagger,~1\leq j\leq L-1,
\end{eqnarray}
where $C_{j\sigma}=c_{j+1,\sigma}-c_{j\sigma},~S_{j\sigma}^\dagger=-c^\dagger_{j+1,\sigma}-c^\dagger_{j\sigma}$. The parent Hamiltonian given in Eqs.~(4) and (6) can then be expanded in position space~(we use the shorthand $\sum_{\boldsymbol k}=\sum_{k_1,k_2,k_3, k_4}$ and $\sum_{\boldsymbol \sigma}=\sum_{\sigma_1,\sigma_2,\sigma_3, \sigma_4}$)
\allowdisplaybreaks
\begin{eqnarray}\label{eq:dbwireHpp_S}
H&=&\sum_{p_1p_2p_3p_4}H_{p_1 p_2;p_3 p_4}\hat{A}^\dagger_{p_1 p_2} \hat{A}_{p_3 p_4}=\frac{16}{L^2}\sum_{\boldsymbol{k},\boldsymbol{\sigma}}\sum^{L}_{j=1}\sin k_1j\sin k_2j\sin k_3j\sin k_4j\nonumber\\
&&{}\times\left[p\delta_{\sigma_1\sigma_2\sigma_3\sigma_4}+q\delta_{\sigma_1\sigma_2}\delta_{\sigma_3\sigma_4}-r(\delta_{\sigma_1\sigma_3}\delta_{\sigma_2\sigma_4}+\delta_{\sigma_1\sigma_4}\delta_{\sigma_2\sigma_3})\right] (C^\dagger_{k_1\sigma_1}S_{k_2 \sigma_2}+C^\dagger_{k_2 \sigma_2}S_{k_1 \sigma_1})(S^\dagger_{k_3\sigma_3}C_{k_4 \sigma_4}+S^\dagger_{k_4 \sigma_4}C_{k_3 \sigma_3})\nonumber\\
&=&\frac{64}{L^2}\sum_{\boldsymbol{k},\boldsymbol{\sigma}}\sum^{L}_{j=1}(C^\dagger_{k_1\sigma_1}\sin k_1j) (S_{k_2 \sigma_2}\sin k_2j) (S^\dagger_{k_3\sigma_3}\sin k_3j) (C_{k_4 \sigma_4}\sin k_4j) 
\nonumber \\ &&{}\times 
\left[p\delta_{\sigma_1\sigma_2\sigma_3\sigma_4}+q\delta_{\sigma_1\sigma_2}\delta_{\sigma_3\sigma_4}-r(\delta_{\sigma_1\sigma_3}\delta_{\sigma_2\sigma_4}+\delta_{\sigma_1\sigma_4}\delta_{\sigma_2\sigma_3})\right]\nonumber\\ 
&=&\sum_{\boldsymbol{\sigma}}\sum^{L}_{j=1}C^\dagger_{j\sigma_1}S_{j\sigma_2} S^\dagger_{j\sigma_3}C_{j \sigma_4} 
\left[p\delta_{\sigma_1\sigma_2\sigma_3\sigma_4}+q\delta_{\sigma_1\sigma_2}\delta_{\sigma_3\sigma_4}-r(\delta_{\sigma_1\sigma_3}\delta_{\sigma_2\sigma_4}+\delta_{\sigma_1\sigma_4}\delta_{\sigma_2\sigma_3})\right]\nonumber\\
&=&(2t-\beta-\gamma)\sum^{L}_{j=1}[C^\dagger_{aj}S_{aj} S^\dagger_{aj}C_{aj} +(a\to b)]+(\gamma-\beta)\sum^{L}_{j=1}[C^\dagger_{aj}S_{aj} S^\dagger_{bj}C_{bj}+(a\leftrightarrow b)]\nonumber\\
&&+(\beta+\gamma)\sum^{L}_{j=1}(C^\dagger_{aj}S_{bj}+C^\dagger_{bj}S_{aj})(S^\dagger_{aj}C_{bj}+S^\dagger_{bj}C_{aj}),
\end{eqnarray}
%\sum_{\sigma_1 \sigma_2 \sigma_3 \sigma_4}\sum_{k_1 k_2 k_3 k_4}
where $\beta=-(q+r)/2,~\gamma=(q-r)/2$ and $t=(p+q-3r)/2$. By expanding the last line of Eq.~(\ref{eq:dbwireHpp_S}) we get the form of Eq.~(\ref{eq:dbwireH}) in the main text~(with $\alpha=r=-\beta-\gamma$).
%$p=2(t+\alpha-\gamma),~q=\gamma-\beta,r=\alpha$
\subsection{Positive region of $\hat{H}$}
We now prove that the matrix $H_{p_1 p_2;p_3 p_4}$ given in Eq.~(\ref{eq:dbwireHpp}) is positive semi-definite in the triangle region shown in Fig.~\ref{fig:triangle} in the main text. Notice that $H_{p_1 p_2;p_3 p_4}=H_{k_1 k_2;k_3 k_4}\cdot \lambda_{\sigma_1\sigma_2;\sigma_3\sigma_4}$ with orbital part $H_{k_1 k_2;k_3 k_4}=\frac{16}{L^2}\sum^{L}_{j=1}\sin k_1j\sin k_2j\sin k_3j\sin k_4j$ and spin part $\lambda_{\sigma_1\sigma_2;\sigma_3\sigma_4}=p\delta_{\sigma_1\sigma_2\sigma_3\sigma_4}+q\delta_{\sigma_1\sigma_2}\delta_{\sigma_3\sigma_4}-r(\delta_{\sigma_1\sigma_3}\delta_{\sigma_2\sigma_4}+\delta_{\sigma_1\sigma_4}\delta_{\sigma_2\sigma_3})$. The orbital part is always positive, since for any vector $f_{k k'}$ we have 
\begin{equation}
\sum_{k_1k_2k_3k_4}f^*_{k_1 k_2}H_{k_1k_2;k_3k_4}f_{k_3k_4}=\frac{16}{L^2}\sum_{k_1k_2k_3k_4}\sum^{L}_{j=1}f^*_{k_1 k_2}\sin k_1j\sin k_2j\sin k_3j\sin k_4jf_{k_3k_4}=\frac{16}{L^2}\sum^{L}_{j=1}f^*_{j}f_{j}\geq 0,
\end{equation}
where $f_j=\sum_{kk'}f_{kk'}\sin kj\sin k'j $. For the spin part $\lambda_{\sigma_1\sigma_2;\sigma_3\sigma_4}$, we write it in the matrix form~(assume the order $aa,ab,ba,bb$)

\begin{eqnarray}
\lambda=
\begin{bmatrix}
p+q-2r & 0 & 0 & q\\
0 & -r & -r & 0\\
0 & -r & -r & 0\\
q & 0 & 0 & p+q-2r
\end{bmatrix}=
\begin{bmatrix}
2t-\gamma-\beta & 0 & 0 & \gamma-\beta\\
0 & -\alpha & -\alpha & 0\\
0 & -\alpha & -\alpha & 0\\
\gamma-\beta & 0 & 0 & 2t-\gamma-\beta
\end{bmatrix}=\lambda_1\oplus\lambda_2,
\end{eqnarray}
with $\lambda_1=\begin{bmatrix}-\alpha & -\alpha \\ -\alpha & -\alpha \end{bmatrix}$ acting on ($ab,ba$) and $\lambda_2=\begin{bmatrix}2t-\gamma-\beta  & \gamma-\beta \\ -\gamma-\beta & 2t-\gamma-\beta  \end{bmatrix}$ acting on $(aa,bb)$. Thus $\lambda$ is positive-definite if and only if both $\lambda_1$ and $\lambda_2$ are positive-definite. The condition that $\lambda_1$ is positive-definite gives $-\alpha>0$, while $\lambda_2$ is positive-definite gives $|\gamma-\beta|<2t-\gamma-\beta$, which simplifies to $\alpha<0,~\beta<t,~\gamma<t$, leading to the triangle region in the main text. 

\subsection{Alternative derivations of the parent Hamiltonian}
The derivation of $\hat{H}$ presented above enables us to see how the double wire model follows from our general construction and can be generalized to arbitrary points of Kitaev's model. However, for the double wire parent Hamiltonian constructed in our main text, simpler derivations exist. Actually, Eq.~(\ref{eq:S_j}) gives us annihilators of the double wire ground state $|G_A\rangle\otimes|G_B\rangle$ in position space. Thus we can directly build the parent Hamiltonian $\hat{H}$ using the real space version of Eqs.~(\ref{eq:Hparen}) and (\ref{eq:A_pp2}) with $C_{j\sigma}=c_{j+1,\sigma}-c_{j\sigma},~S_{j\sigma}^\dagger=-c^\dagger_{j+1,\sigma}-c^\dagger_{j\sigma}$~(or equivalently, directly go to the last line of Eq.~(\ref{eq:dbwireHpp_S}) without working in momentum space at all), leading to the same Hamiltonian Eq.~(\ref{eq:dbwireH}) in our main text. This derivation is a direct generalization of the one given in Ref.~\cite{SDiehl}.

Another simple derivation is based on using an alternative basis of single wire ground states~(at $\Delta=t,\mu=0$)~\cite{AoP2014}
\begin{eqnarray}
|G^\eta\rangle&=&(1+\eta c^\dagger_1)(1+\eta c^\dagger_2)\cdots(1+\eta c^\dagger_L)|0\rangle,~~\eta=\pm 1,
\end{eqnarray}
and observing the following properties
\begin{eqnarray}
c^\dagger_ic^\dagger_{i+1}|G^\eta\rangle&=&n_i n_{i+1}|G^\eta\rangle,~~
c_{i+1}c_i|G^\eta\rangle=\bar{n}_i \bar{n}_{i+1}|G^\eta\rangle,\nonumber\\
c^\dagger_ic_{i+1}|G^\eta\rangle&=&n_i \bar{n}_{i+1}|G^\eta\rangle,~~
c^\dagger_{i+1}c_i|G^\eta\rangle=\bar{n}_i n_{i+1}|G^\eta\rangle,
\end{eqnarray} 
where $\bar{n}_i\equiv 1-n_i$. With this, it is easy to check that the number-conserving single wire operator $[(c^\dagger_jc_{j+1}+\mathrm{H.c.})+2(n_j-\frac{1}{2})(n_{j+1}-\frac{1}{2})-\frac{1}{2}]$ annihilates $|G^\eta\rangle$. To include interwire couplings, we notice that
\begin{eqnarray}
J^\dagger_{||,j}J_{||,j}|G^{\eta_A,\eta_B}\rangle=J^\dagger_{=,j} J_{=,j}|G^{\eta_A,\eta_B}\rangle= J^\dagger_{\times,j} J_{\times,j}|G^{\eta_A,\eta_B}\rangle=(U^\square_{j}-U^{(3p)}_{j})|G^{\eta_A,\eta_B}\rangle,
\end{eqnarray}
where $\hat{U}^{\square}_{j}=n^a_j n^a_{j+1}n^b_jn^b_{j+1},~~\hat{U}^{(3p)}_j=[n^a_j n^a_{j+1}(n^b_j+n^b_{j+1})+(a\leftrightarrow b)]$, and $|G^{\eta_A,\eta_B}\rangle=|G^{\eta_A}_A\rangle\otimes|G^{\eta_B}_B\rangle$ is the double wire ground state constructed by direct product of single wire ground states. Therefore
\begin{equation}
(\alpha J^\dagger_{||,j}J_{||,j}+\beta J^\dagger_{=,j} J_{=,j}+\gamma J^\dagger_{\times,j} J_{\times,j})|G^{\eta_A,\eta_B}\rangle=0,
\end{equation}
for $\alpha+\beta+\gamma=0$. It then follows that the Hamiltonian constructed in Eq.~(\ref{eq:dbwireH}) in the main text satisfies $\hat{H}|G^{\eta_A,\eta_B}\rangle=0$, i.e. $|G^{\eta_A,\eta_B}\rangle$ is an eigenstate of $\hat{H}$ with zero energy. This method can only tell us that $|G^{\eta_A,\eta_B}\rangle$ is an eigenstate of $\hat{H}$. To find the positive region of $\hat{H}$~(where $|G^{\eta_A,\eta_B}\rangle$ become its ground states), we still have to turn to other means. 

\begin{figure}
\includegraphics[width=0.5\linewidth]{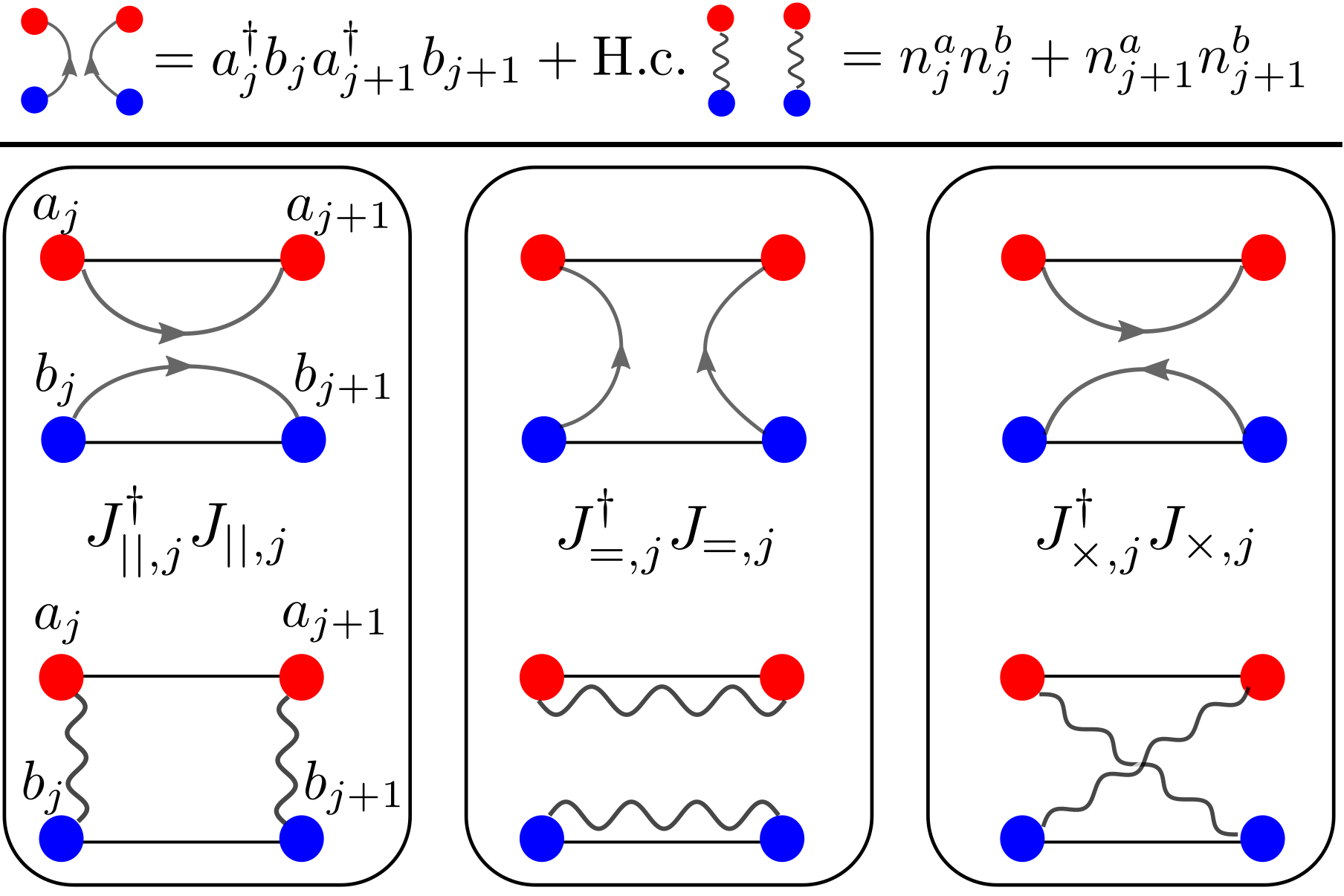}
\caption{\label{fig:diag} Diagrammatic representation of interaction terms in the double wire parent Hamiltonian in Eq.~(\ref{eq:dbwireH}). Double arrows represent pair hopping while wavy lines represent interactions. }
\end{figure}

In Fig.~\ref{fig:diag} we draw a pictorial representation of the interaction terms of the parent Hamiltonian Eq.~(\ref{eq:dbwireH}), including two types of nonlinear terms: interactions and correlated pair tunnelings. 

\end{document}